\documentclass[aip,apl,reprint]{revtex4-1}
\usepackage{amssymb,amsmath}
\usepackage{pgf}
\usepackage{color}
\usepackage{epstopdf}
\usepackage{graphicx}% Include figure files
\usepackage{dcolumn}% Align table columns on decimal point
\usepackage{bm}% bold math
\usepackage[mathlines]{lineno}% Enable numbering of text and display math
\begin{document}

\preprint{AIP/APL}

\title[]{Confined one-way mode at magnetic domain wall for broadband high-efficiency one-way waveguide, splitter and bender}

\author{Xiaogang Zhang}
\affiliation{State Key Laboratory of Functional Materials for Informatics, Shanghai Institute
of Microsystem and Information Technology, Chinese Academy of Sciences, Shanghai 200050,
China}
\affiliation{Graduate
School of Chinese Academy of Sciences, Beijing 100049, People's
Republic of China}
\author{Wei Li}
\email{waylee@mail.sim.ac.cn}
\author{Xunya Jiang}%
 \email{xyjiang@mail.sim.ac.cn}
\affiliation{State Key Laboratory of Functional Materials for Informatics, Shanghai Institute
of Microsystem and Information Technology, Chinese Academy of Sciences, Shanghai 200050,
China}%

\begin{abstract}
We find the one-way mode can be well-confined at the magnetic domain wall by the
photonic bandgap of gyromagnetic bulk material. Utilizing the well-confined one-way mode at the domain wall, we demonstrate the photonic one-way waveguide, splitter and bender can be realized with simple structures, which are predicted to be high-efficiency, broadband, frequency-independent, reflection-free, crosstalk-proof and robustness against disorder. Additionally, we find that the splitter and bender in our proposal can be transformed into each other with magnetic control, which may have great potential applications in all photonic integrated circuit.

\end{abstract}

%\pacs{85.70.Sq, 41.20.Jb, 42.70.Qs, 42.25.Bs }% PACS, the Physics and Astronomy
                             % Classification Scheme.
%\keywords{one-way waveguide, splitter, bender, one-way mode, magnetic domain wall}%Use showkeys
%class option if keyword
                              %display desired
\maketitle

%\section{Introduction}
One-way waveguide has been drawn much attention in recent years \cite{PhysRevA.78.033834,PhysRevA.78.023804,PhysRevLett.100.013904,yu:121133,yu:171116,10.1063/1.3358386,nature08293,PhysRevLett.100.013905,PhysRevLett.100.023902,PhysRevB.84.045425,PhysRevB.50.11187},
for their great potential application in all photonic integrated
circuit. A popular solution for one-way waveguide is utilizing the
gyromagnetic photonic crystal\cite{nature08293,PhysRevLett.100.013905}, which can support a chiral
edge state that exhibits an anomalous unique directionality. Due to the unidirectionality of chiral
edge state, the one-way waveguide can be realized via confinement of the surface states\cite{PhysRevLett.100.013905,nature08293}. Such one-way photonic crystal waveguides with chiral
edge state has been proven to exhibit strong robustness against disorder\cite{PhysRevLett.100.023902,PhysRevB.84.045425}. However, the one-way photonic crystal waveguides need complex structures, which may bring more challenges in integrated circuits. In addition, the existing one-way photonic crystal waveguides are all frequency-sensitive. To avoid these difficulties, very recently, another solution for one-way propagation has been proposed\cite{Zhu:10}, in
which a broadband one-way mode (OWM) is predicted to propagate along a magnetic domain wall. However, the cost of the solution is the broadband OWM may be not suitable for one-way waveguide, since the mode extends into the bulk rather than
confinement at the domain wall.

In reviewing these existing efforts, we feel desirable to find a design for one-way waveguide that should include at
least three characteristics: (I) simple structure; (II) broadband working frequency; and (III)
well-confined mode in the waveguide. In this Letter, we will present our design to serve this purpose.

Our design is based on the domain wall. The key of this work is to find a confinement
mechanism to localize the broadband OWM at the domain wall. We find the OWM can be confined well
at the domain wall via the \emph{photonic bandgap} of bulk material \cite{xgzh:2011,230282581}, unlike the one-way photonic crystal waveguides\cite{nature08293,PhysRevLett.100.013905} confining the OWM in waveguide by the photonic bandgap of gyromagnetic photonic crystal. %although the origin of the photonic bandgap of our structure is different from that of GPC\cite{nature08293,PhysRevLett.100.013905}.

Utilizing the OWM well-confined at the domain wall, an one-way waveguide with a simple structure and broadband working
frequency can be realized. Such OWM makes the one-way waveguide high-efficiency and very robust against disorder.
Besides one-way waveguide, the OWM localized at domain wall can also be used to design high-efficiency, crosstalk-proof
splitters and benders. Additionally, with magnetic controlling, the splitter and bender can be transformed into each other in our proposal, which may have great potential application in all photonic integrated circuit.

\begin{figure}%[H]
\centering
\includegraphics[width=1.1\columnwidth]{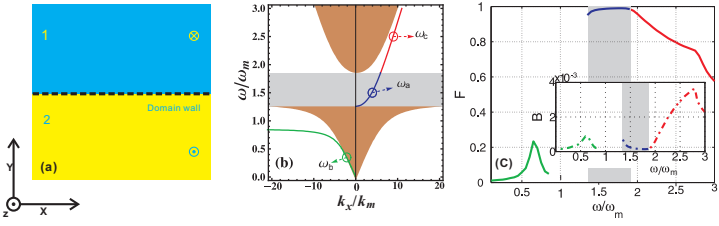}
\caption{\label{fig:fig1} (a) The model of domain wall. The external dc magnetic
field for $y>0$ (blue area) and $y<0$ (yellow area) are applied along $-z$ and $+z$, respectively. (b)The dispersion relation of TE mode and the projected band structure of bulk modes. The gray and the brown regions represent the photonic bandgap and the extended modes of the bulk respectively. Different OWMs including positive propagation OWM-I, positive propagation OWM-II and negative propagation OWM-II are represented by the blue one, the red one and the green one, respectively. (c) The forward transmission $F$ (solid lines) and backward transmission $B$ (dot-dashed lines in the inset box) of OWM-I and OWM-II versus frequency. The distance between the detector and the source is $100mm$.}
\end{figure}

%\section{Model and Analysis}
Our model for OWM is schematically shown in Fig.\ref{fig:fig1}(a), in which a domain wall exists at the interface between two gyromagnetic media such as yttrium-iron-garnet(YIG)\cite{PhysRevB.78.155101} at $x$-$y$ plane with anti-parallel dc magnetic fields along $\pm z$. We use blue area and yellow area to denote $-z$ and $+z$ dc magnetic fields respectively in this figure and subsequent figures. According to real YIG material, the relative permittivity $\epsilon=15$ and relative permeability would has a gyromagnetic form\cite{PhysRevB.78.155101}:
\begin{equation}\label{permeability}
\overleftrightarrow{\mu}=\left[\begin{array}{ccc}
\mu_1&\pm j\mu_2&0\\
\mp j\mu_2&\mu_1&0\\
0&0&1
\end{array}
 \right]
 \end{equation}
where $\mu _1=1+\frac{\omega_m(\omega _0-i\alpha\omega)}{(\omega _0-i\alpha\omega)^2-\omega^2}$, $\mu _2=\frac{\omega \omega _m}{[(\omega _0-i\alpha\omega)^2-\omega ^2]}$. $\omega_0=\gamma H_0$ is the resonance frequency with $\gamma$ as the gyromagnetic ratio, and $H_0$ the sum of the external dc magnetic field. $\alpha$ is the damping coefficient, and $\omega_m$ is the characteristic circular frequency corresponding to a wavevector $k_m=\omega_m/c$. $\pm$ and $\mp$ describe the direction of $H_0$. In this structure, only TE wave can be formed as an OWM propagating along the domain wall.

In order to illustrate how to confine an OWM at the domain wall via \emph{photonic bandgap}, we study the dispersion relation of bulk modes and OWMs in our structure, by exactly solving the Maxwell equations\cite{xgzh:2011}, and the results are shown in Fig.\ref{fig:fig1}(b). In this figure, the solid lines correspond to the OWMs of domain wall. The brown and gray regions correspond to the projected band structure of bulk modes. For the bulk modes, a gap (gray region) exists obviously. Therefore, we can group the OWMs into two types: OWM-I and OWM-II, whose frequency range are respective in the gap and in the band of bulk modes. Comparison OWM-I with OWM-II, their forward transmission\cite{Zhu:10} $F$ and backward transmission\cite{Zhu:10} $B$ are clearly different, as shown in Fig.\ref{fig:fig1}(c). In this figure, we can see the $F$ and $B$ of OWM-I (in gray region) are almost unit and zero, respectively, and they are nearly frequency-independent; while for OWM-II, the forward transmission becomes smaller and the backward transmission becomes larger, and both are frequency-dependent. These results indicate that OWM-I is a much higher efficient unidirectional mode. In addition, we can also find that OWM-I is broadband, with working frequency range
$1.28\omega_m$ to $1.88\omega_m$ ($\omega_m=5.26GHz$ in this work).

\begin{figure}%[H]
\centering
\includegraphics[width=0.9\columnwidth]{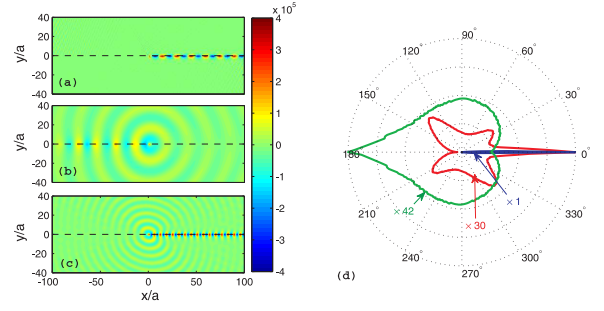}
\caption{\label{fig:fig2} (a-c) Steady-state scattered fields of OWM$_a$, OWM$_b$ and OWM$_c$ scattered by a PEC particle located at the center of domain wall. (d)Radiation pattern. Angular dependence of the energy flux is plotted as a
function of polar angle. The $0^\circ$ and $180^\circ$ mark $+x$ and $-x$ direction, respectively. The blue, green and red lines correspond to the scattered energy flux of OWM$_a$, OWM$_b$ and OWM$_c$, respectively.}
\end{figure}

Physically, the reason that OWM-I exhibits higher efficient unidirectionality than OWM-II can be explained
as follows. Only OWM-I can be localized at the domain wall, since it becomes evanescent in the bulk with
the intensity decay exponentially away from the domain wall. On the contrary, OWM-II is much
easier to extend to the bulk, because their photon energies degenerate with the bulk modes. To show this, three typical OWMs (i.e., OWM$_a$, OWM$_b$ and OWM$_c$ with frequency $\omega_a$=$0.3\omega_m$, $\omega_b$=$1.5\omega_m$ and $\omega_c$=$2.5\omega_m$, respectively) are studied, which are excited by a current source at the domain wall. Obviously, OWM$_a$ belongs to OWM-I, while OWM$_b$ and OWM$_c$ are the type of OWM-II. With a single perfect-electric-conductor (PEC) particle scatter located at the domain wall at a distance $d=2mm$ from the source (the scatter is on the right of the source for OWM$_a$ and OWM$_c$, and on the left for OWM$_b$), we calculate the scattered fields of the three OWMs and the radiation patterns of the scatter by Finite-Differential-Time-Domain (FDTD) method\cite{1138693}, which are shown in Fig.\ref{fig:fig2}. From Fig.\ref{fig:fig2}(a), we can see the scattered field of OWM$_a$ is still well-confined at the domain wall, but others extends into bulks as shown in Fig.\ref{fig:fig2}(b) and (c). The fact that the confinement quality of OWM-I is much better than that of OWM-II can be also illustrated by the radiation pattern of the scatter, as shown in Fig.\ref{fig:fig2}(d). In this figure, we can see that only OWM$_a$ can concentrate almost all energy flux propagating along the domain wall, without reflection and diffraction; while OWM$_b$ and OWM$_c$ have quite a bit of energy flux scattered to other directions. With respect to the directional coefficient\cite{antennaBook} of the three modes along the domain wall ($0^\circ$ for OWM$_a$ and OWM$_c$, $180^\circ$ for OWM$_b$), we find OWM$_a$ is about 42 and 30 times larger than that of OWM$_b$ and OWM$_c$, respectively.

\begin{figure}%[H]
\centering
\includegraphics[width=0.9\columnwidth]{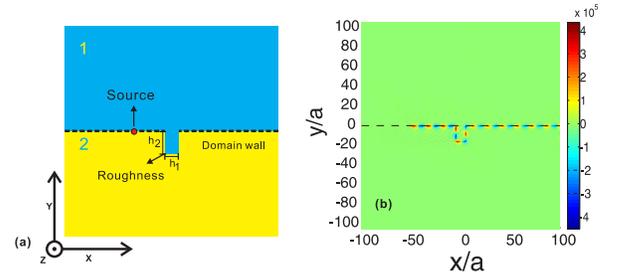}
\caption{\label{fig:fig3} (a) The structure of the domain wall with roughness. (b)
Steady-state $E_z$ field distribution of OWM$_a$.%embedded with a YIG slab defect.
}
\end{figure}

Furthermore, such OWM-I is very robust against disorder, since the disorder-induced backscattering
is suppressed due to its unidirectionality.
This phenomenon is very useful to keep the one-way waveguide high efficiency when the
domain wall is rough. To show this in simplicity, as shown in Fig.\ref{fig:fig3}(a), a
raised portion is chosen to represent the roughness, with $h_1$=$10a$ and $h_2$=$15a$, where $a$=$1mm$ is the length unit in this work. The OWM
of this domain wall with roughness is roused and propagates only along one direction, as shown in
Fig.\ref{fig:fig3}(b).

\begin{figure}%[H]
\centering
\includegraphics[width=0.9\columnwidth]{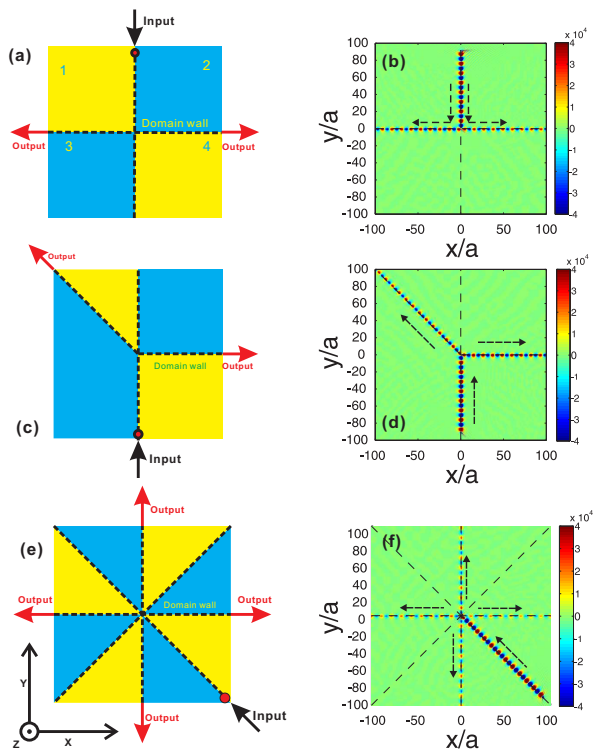}
\caption{\label{fig:fig4} The structures of splitters and their $E_z$ field distribution of OWM$_a$. (a) 50\% beam splitter, and (b) its field distribution. (c) Any-angle beam splitter, and (d) its field distribution. (e) Multiple-beam splitter, and (f) its field distribution. The dashed arrow lines indicate the direction of energy flux. }
\end{figure}

\begin{figure}%[H]
\centering
\includegraphics[width=0.9\columnwidth]{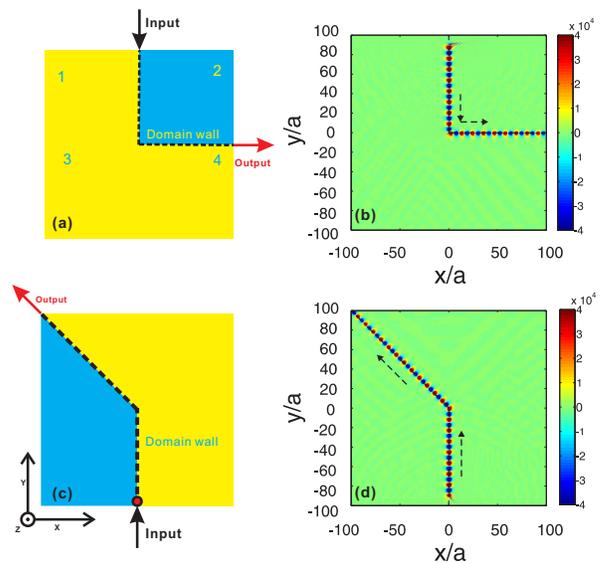}
\caption{\label{fig:fig5} Benders based on domain wall with OWM$_a$. (a) $90^\circ$ bender, and (b) its $E_z$ field distribution. (c) Any-angle bender, and (d) its $E_z$ field distribution of the any-angle bender. The dashed arrow lines indicate the direction of energy flux. }
\end{figure}

%\section{Application}
As discussed above, OWM-I is very suitable to realize a high-efficiency one-way waveguide. Due to the unidirectionality and well confinement of OWM-I at domain wall, it can also be used to design more devices with simple structures in the realm of all photonic integrated circuit, such as splitter and bender, which are predicted to exhibit high-efficiency, broadband, frequency-independent, reflection-free, crosstalk-proof and robustness against disorder. As typical examples, Fig.\ref{fig:fig4} and Fig.\ref{fig:fig5} show our proposal of splitters and benders, respectively. Our simulation results show that all these devices have the same broadband as the one-way waveguide in Fig.\ref{fig:fig1}.

In Fig.\ref{fig:fig4}(a), a one-way cross-domain wall splitter based on the domain wall is constituted by four YIG bulks. Utilizing OWM-I, a beam (OWM$_a$) from the ``input" port is split into two equal intensity beams and exit from two ``output" ports, as shown in Fig.\ref{fig:fig4}(b). It is proved that the splitter is a high-efficiency 50\% splitter, since the beams propagating along the domain wall are unidirectional and reflection-free. Furthermore, the beams at different domain walls are immune from crosstalk, because the OWM-I is well-confined .

Similarly, based on the simple structure of domain wall with OWM-I, other one-way cross-domain wall splitters can be designed. For instance, an any-angle splitter and a multiple-beam splitter with simple structures, are presented in Fig.\ref{fig:fig4}(c) and Fig.\ref{fig:fig4}(e), respectively. And their field distributions are shown in Fig.\ref{fig:fig4}(d) and Fig.\ref{fig:fig4}(f), respectively. From the field distributions, we can find that all the splitters exhibit high-efficiency, without reflection, diffraction and crosstalk.

Based upon OWM-I, we find another important application such as a sharp bender. In Fig.\ref{fig:fig5}(a) and Fig.\ref{fig:fig5}(c), a $90^\circ$ bender and an any-angle splitter are illustrated, respectively. Due to OWM-I, both benders are high-efficiency, which can be seen from the field distributions as shown in Fig.\ref{fig:fig5}(b) and (d). Actually, the $90^\circ$ bender can be transformed from the splitter in Fig.\ref{fig:fig4}(a), just by reversing the dc magnetic field direction of area-3 in Fig.\ref{fig:fig4}(a), and, the reverse process is also feasible. Following this method, the any-angle bender in Fig.\ref{fig:fig5}(c) and the any-angle splitter in Fig.\ref{fig:fig4}(c) can also be transformed into each other. The realization of this transformation requires some magnetic control techniques\cite{nature08293,nm1711} such as homogenization and exact location technique. Therefore, with magnetic control, there is a reversible transformation between splitters and benders in our proposal, which may have great potential applications in all photonic integrated circuit.

This work was supported by the NSFC (Grant Nos. 11004212, 11174309, 60877067 and 60938004), and the STCSM (Grant Nos. 11ZR1443800 and 11JC1414500).

\nocite{*}
%\bibliography{aipsamp}% Produces the bibliography via BibTeX.
\bibliography{splitter}
\end{document}